\documentclass[twocolumn,aps,prl,groupedaddress,showpacs]{revtex4}
\usepackage{epsfig,amssymb}
\def\comment#1{}\def\labell#1{\label{#1}}
\begin{document}
\title{Entanglement assisted capacity of the broadband lossy channel}
\author{Vittorio Giovannetti$^1$, Seth Lloyd$^{1,2}$, Lorenzo
  Maccone$^1$, and Peter W. Shor$^3$} \affiliation{Massachusetts
  Institute of Technology -- $^1$Research Laboratory of Electronics,
  $^2$Department of Mechanical Engineering\\ 77 Massachusetts Ave.,
  Cambridge, MA 02139, USA\\ $^3$\mbox{AT\&T Labs - Research 180 Park
    Ave, Florham Park, NJ 07932, USA}} \date{\today}

\begin{abstract}
  We calculate the entanglement assisted capacity of a multimode
  bosonic channel with loss. As long as the efficiency of the channel
  is above $50\%$, the superdense coding effect can be used to
  transmit more bits than those that can be stored in the message sent
  down the channel.  Bounds for the other capacities of the multimode
  channel are also provided.
\end{abstract}
\maketitle

Among the zoology of different capacities of quantum channels
{\cite{shor,chuang}}, the entanglement assisted classical capacity
$C_E$ plays an important role. This quantity has been introduced in
{\cite{shorprl}} to measure the amount of classical information that
can be sent through the channel in the presence of an unlimited
quantity of prior entanglement between sender and receiver. $C_E$ and
its quantum counterpart $Q_E=C_E/2$ (i.e.  the amount of qubits that
can be sent in the presence of an unlimited quantity of prior
entanglement) give upper bounds to the classical and quantum
capacities of the channel, including the unassisted capacities whose
values are yet to be determined.  Moreover, it has been conjectured
{\cite{shor2}} that the entanglement assisted classical capacity
defines a class of equivalences since all channels with the same $C_E$
seem to be able to efficiently simulate one another. Unlike the case
of most of the other capacities, it has a closed expression in terms
of the quantum mutual information
\begin{eqnarray}
I({\cal N},\varrho)=S(\varrho)+S({\cal N}[\varrho])-S(({\cal
N}\otimes\openone)[\Phi_\varrho])
\;\labell{qmi},
\end{eqnarray}
where $S(\varrho)=-$Tr$[\varrho\log_2\varrho]$ is the Von Neumann
entropy, $\cal N$ is the map that describes the communication channel
and $\Phi_\varrho$ is a purification of the input density matrix
$\varrho$. The value of $C_E$ is the maximum of $I({\cal N},\varrho)$
over all the possible inputs $\varrho$ to the channel
{\cite{shor2,holevoce}}.

The entanglement assisted capacity for bosonic Gaussian channels was
analyzed in {\cite{holevo}}, where it was shown that the maximization
in the expression of $C_E$ can be performed over Gaussian states.
These channels are important because they are the basic building
blocks of bosonic communication schemes and because they allow one to
describe infinite dimensional systems with techniques from finite
dimensional linear algebra. In this paper we derive $C_E$ for
multimode bosonic channels in the presence of loss and average input
energy constraint, and use these results and the techniques developed
to provide upper and lower bounds for other channel capacities. We
calculate $C_E$ for the multimode channel as the sum of the
entanglement assisted capacities of the single modes maximized over
non-squeezed Gaussian states. In fact, $C_E$ is additive and we show
that squeezing the input states does not increase the $C_E$ of a
single mode.  For generic values of the channel quantum efficiency
$\eta$ we cannot provide an analytical expression for $C_E$, but we
give a general characterization and a numerical solution. For
$\eta=1/2$, the value of $C_E$ can be analytically solved and,
interestingly, shown to coincide with the wideband lossless channel
capacity {\cite{caves}}.

{\em Broadband lossy channel.---}
In the Heisenberg picture the $i$th mode of the lossy channel with
quantum efficiency $\eta_i$ evolves as
\begin{eqnarray}
a'_i=\sqrt{\eta_i}\;a_i +\sqrt{1-\eta_i}\;b_i
\;\labell{ch},
\end{eqnarray}
where $a_i$, $a'_i$ and $b_i$ are the annihilation operators of the
input, output and noise modes respectively.  The loss map ${\cal N}_i$
for the $i$th mode arises by tracing away the noise mode $b_i$ (in the
vacuum state) and the global loss map $\cal N$ is the tensor product~
$\bigotimes_i{\cal N}_i$. The channel described by $\cal N$ maps
Gaussian input states into Gaussian output states and is hence a
Gaussian channel.

The calculation of $C_E$ for the multimode lossy channel stems from
the following three facts: {\it i)}~the additivity property of the
entanglement assisted capacity, from which the $C_E$ of the channel is
calculated as the sum of the $C_E$ of each mode
{\cite{cerfadami,shor}}, i.e.
\begin{eqnarray} 
C_E=\max_{\varrho_{j}\in{\cal H}_{j}}
\Big\{\sum_{i}I({\cal N}_i,\varrho_i)\Big\} 
\;\labell{addit},
\end{eqnarray}
where ${\cal H}_j$ is the Hilbert space of the $j$th mode of the
channel, and the max is taken over the states $\varrho_i$ that satisfy
the average energy constraint
\begin{eqnarray}
\sum_{i}\hbar\omega_iN_i={\cal E}
\;\labell{energ},
\end{eqnarray}
with $\omega_i$ the frequency of the $i$th mode and $N_i$ its average
number of photons; {\it ii)}~the Holevo-Werner theorem according to
which the maximum of $I({\cal N}_i,\varrho_i)$ for Gaussian channels
can be evaluated on Gaussian input states {\cite{holevo}}; {\it
  iii)}~the fact that squeezing the input does not increase $C_E$, so
that it can be estimated on non-squeezed inputs: as shown in the
appendix, the maximum value of $I({\cal N}_i,\varrho_i)$ (fixing the
energy in the $i$th mode) is obtained when $\varrho_i$ does not
contain any squeezing and is given by
\begin{eqnarray}
c_E(N_i,\eta_i)=g(N_i)+g(\eta_i N_i)-g((1-\eta_i)N_i)
\;\labell{cedii},
\end{eqnarray}
where the function $g$ is defined as
\begin{eqnarray}
g(x)\equiv (x+1)\log_2(x+1)-x\log_2(x)
\;\labell{defdig},
\end{eqnarray}
for $x\neq 0$ and $g(0)=0$.  The total entanglement assisted capacity
is then\begin{eqnarray} C_E=\max_{N_j}\sum_{i}c_E(N_i,\eta_i)
  \;\labell{cdie},
\end{eqnarray}
where the maximum is taken over the sets $\{N_j\}$ satisfying the
energy constraint (\ref{energ}).

The maximization (\ref{cdie}) can be performed using the Lagrange
multiplier procedure, which, for
$\eta\neq 0,1$, gives the following equation {\cite{notaeta}}
\begin{eqnarray} &&\left(1+\frac 1{N_j}\right)\left(1+\frac
1{\eta_j\;N_j}\right)^{\eta_j}\nonumber\\&&
\qquad\qquad=e^{\omega_j/\Omega}\left(1+\frac
1{(1-\eta_j)N_j}\right)^{1-\eta_j}
\;\labell{lagr1},
\end{eqnarray}
where $1/(\Omega\ln 2)$ is the Lagrange multiplier that must be chosen
to satisfy the constraint (\ref{energ}). In general this equation is
difficult to solve analytically, but we can still give some
characterization of the solution, at least when all the quantum
efficiencies coincide (i.e.  $\eta_j=\eta$ for all $j$). In this case
the solution of Eq.~(\ref{lagr1}) is a function of $\omega_j/\Omega$
and $\eta$, i.e. $N_j={\cal F}(\omega_j/\Omega,\eta)$. To derive
$\Omega$ we use Eq.~(\ref{energ}) that becomes
\begin{eqnarray}
\frac{\cal E}\hbar=\sum_{i}\omega_i{\cal F}(\omega_i/\Omega,\eta)
\simeq\int_0^\infty \frac{d\omega}{\delta\omega}\;\omega\;{\cal
F}(\omega/\Omega,\eta)
\;\labell{condiz},
\end{eqnarray}
where we have replaced the sum over the mode index $i$ with an
integral over the mode frequencies, assuming that the minimum
frequency interval $\delta\omega$ of the channel is small. With a
variable change in the integral (\ref{condiz}), we find that
$\Omega=\sqrt{2\pi\; {\cal P}/[f(\eta)\;\hbar]}$ where ${\cal P}={\cal
  E}\delta\omega/(2\pi)$ is the wideband channel input power during
the transmission time ${\cal T}=2\pi/\delta\omega$ and
\begin{eqnarray} f(\eta)\equiv\int_0^\infty dx\;x\;{\cal F}(x,\eta)
\;\labell{defdif}.
\end{eqnarray}
The value of $C_E$ is then obtained placing the solution of
Eq.~(\ref{lagr1}) to evaluate the sum (\ref{cdie}), i.e.
\begin{eqnarray} C_E\simeq\int_0^\infty
  \frac{d\omega}{\delta\omega}\;c_E({\cal F}(\omega/\Omega,\eta),\eta)
  \;\labell{cidie1}.
\end{eqnarray}
Performing again a change of integration variables, we finally find
\begin{eqnarray} C_E={\cal T}\;\frac{1}{\ln 2}\sqrt{\frac{\pi\;{\cal
P}}{3\;\hbar}}\;{\cal C}(\eta)
\;\labell{cidiefin},
\end{eqnarray}
where     \begin{eqnarray} {\cal C}(\eta)\equiv\frac {\ln
2}{\pi}\sqrt{\frac 3{2\;f(\eta)}}\int_0^\infty\!\!  dx\;c_E({\cal
F}(x,\eta),\eta) \;\labell{calc}.
\end{eqnarray}
Notice that, even without knowing the explicit form of the function
${\cal C}(\eta)$, Eq.~(\ref{cidiefin}) gives the exact dependence on
the input power of the entanglement assisted capacity for the channel
{\cite{nota1}}. In particular, the entanglement assisted capacity per
unit time of channel use $R_E\equiv C_E/{\cal T}$ is proportional to
the rate $R_C=\frac 1{\ln 2}\sqrt{\frac{\pi\;{\cal P}}{3\;\hbar}}$ of the
wideband noiseless bosonic channel without prior entanglement
{\cite{caves}}, i.e.  $R_E=R_C\;{\cal C}(\eta)$.

{\em General properties of $C_E$.---} The form of ${\cal C}(\eta)$ is
not easily determined analytically, but we can still calculate it for
some values of $\eta$. First of all, for $\eta=0$ all the
$c_E(N_i,\eta)$ are null and ${\cal C}(0)=0$: no photons arrive, and
no bits are transferred.  Interestingly, for $\eta=1/2$
Eq.~(\ref{lagr1}) can be solved analytically and has solution
\begin{eqnarray}
N_j=\frac 1{e^{\omega_j/\Omega}-1}
\;\labell{soluz}.
\end{eqnarray}
In this case, $f(1/2)={\pi^2}/6$ and ${\cal C}(1/2)=1$, and hence the
entanglement assisted capacity for the $\eta=1/2$ wideband channel
equals the unassisted capacity of the noiseless wideband channel
${\cal T}R_C$ {\cite{caves}}: prior entanglement is sufficient to
restore perfect transmission for a 50$\%$ lossy channel (this result
holds also for the single mode channel--- see appendix). The solution
can be linearized around $\eta=1/2$ and the first order Taylor
expansion of ${\cal C}(\eta)$ can be obtained as
\begin{eqnarray} {\cal C}(\eta)=\frac 32\Big(\eta-\frac
12\Big)+1+{\cal O}((\eta-1/2)^2)
\;\labell{linearizz}.
\end{eqnarray}
The case $\eta=1$ can be completely solved too, given that the
Lagrange equation has the same solution (\ref{soluz}) of the case
$\eta=1/2$. Here, since $c_E(N_i,1)=2c_E(N_i,1/2)$, we find   ${\cal
C}(1)=2{\cal C}(1/2)=2$: the entanglement assisted capacity for the
noiseless channel is twice the unassisted capacity as predicted by the
superdense coding effect {\cite{sdc}}. In Fig.~\ref{f:numeric}a   ${\cal
C}(\eta)$ is numerically evaluated and plotted along with the
linearization (\ref{linearizz}).  The fact that ${\cal C}(\eta)>1$ for
$\eta>1/2$ shows that, even in the presence of noise, prior
entanglement allows one to transmit more bits than those actually sent
in the channel (i.e. ${\cal T}R_C$) thanks again to the superdense coding
effect. A similar effect has been shown also for the erasure channel
{\cite{erasure,shorprl}}.

An interesting class of lower bounds, that provides a good analytical
approximation for $C_E$ can be obtained by considering the set
(parametrized by $\zeta>0$)
\begin{eqnarray} 
N_j=\frac{\zeta^2}{e^{\zeta\omega_j/\Omega_0}-1}
\;\labell{ssss},
\end{eqnarray}
where $\Omega_0={6\ln 2}R_C/\pi$.  Using Eq.~(\ref{ssss}), we find the
bound
\begin{eqnarray}
{\cal
C}(\eta)\geqslant[{\Lambda(\zeta^2)+\Lambda(\eta\zeta^2)-
\Lambda((1-\eta)\zeta^2)}]/[{\zeta\Lambda(1)}]
\;\labell{ss1},
\end{eqnarray}
where   $\Lambda(y)\equiv\int_0^\infty dx\;g\left(\frac
y{e^x-1}\right)$.  In particular, the case $\zeta=1$ (see
Fig.~\ref{f:numeric}a) corresponds to employing the exact solution for
$\eta=1/2,1$ of Eq.~(\ref{soluz}) for any value of $\eta$.

\begin{figure}[hbt]
\begin{center}
\epsfxsize=.93
\hsize\leavevmode\epsffile{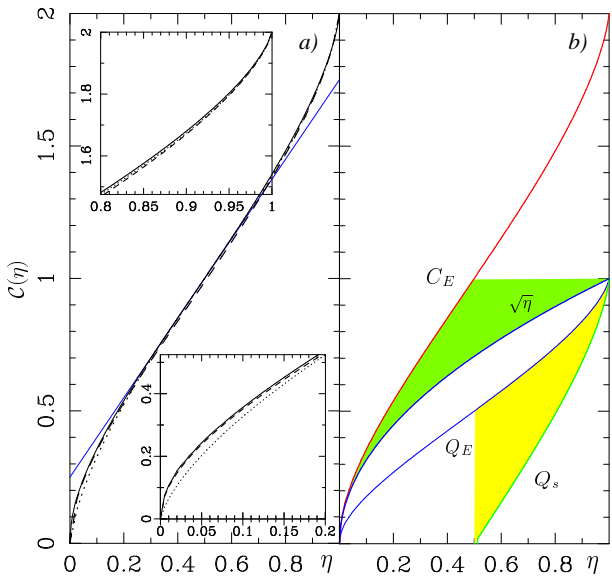}
\hskip .3cm$\;$
\end{center}
\vspace{-.5cm}
\caption{{\it a)}~Plot, as function of the quantum efficiency $\eta$,
  of the numerical solution for ${\cal C}(\eta)$ of Eq.~(\ref{calc})
  (continuous line), of the linearization (\ref{linearizz}) (gray
  line), and of the lower bounds (\ref{ss1}) with $\zeta=1$ (dotted
  line) and $\zeta=1/\sqrt{\eta}$ (dashed line). The inserts show the
  same graph in the regions of small and large $\eta$. The points
  above ${\cal C}(\eta)=1$ (i.e. for $\eta>1/2$) show where the
  superdense coding effect allows a lossy channel to beat the capacity
  of the noiseless channel without prior entanglement. {\it
    b)}~Classical and quantum capacities of the lossy wideband
  channel. The classical capacity $C/({\cal T}R_C)$ is confined in the
  dark gray area between the upper bound given by $C_E$ and the lower
  bound $\sqrt{\eta}$. The quantum capacity $Q/({\cal T}R_C)$ is
  confined in the light gray area between the upper bound given by the
  entanglement assisted quantum capacity $Q_E=C_E/2$ and the lower
  bound $Q_s$ obtained by calculating the coherent information
  according to Eq.~(\ref{coh}). $Q$ is null for $\eta\leqslant 1/2$.}
\labell{f:numeric}\end{figure}

{\em Capacity bounds.---}
The classical capacity $C$ and the quantum capacity $Q$ measure
respectively the number of bits and qubits that can be sent reliably
through the channel per channel use (without the aid of prior
entanglement).  Unlike the case of $C_E$, for $\eta\neq 1$ a closed
expression for $C$ is not known nor it is known whether this quantity
is additive {\cite{shor}}: it may be that entangling successive uses
of the channel one can increase the amount of information transmitted.
Limiting the analysis to unentangled coding procedures, a lower bound
for $C$ can be obtained as {\cite{schumi1}}
\begin{eqnarray}
C\geqslant\max_{p_j{(\mu)},\rho_j{(\mu)}}\sum_{i}{\cal
X}\left(p_i{(\mu)},\rho_i{(\mu)}\right)\labell{add}\;,
\end{eqnarray}
where $\varrho_i=\int d\mu\;p_i{(\mu)}\rho_i{(\mu)}$ describes a
message in which the ``$\mu$th letter'' $\rho_i{(\mu)}$ in the $i$th
mode has probability density $p_i{(\mu)}$ and where ${\cal X}$ is the
Holevo information $S({\cal N}_i[\varrho_i])-\int
d\mu\;p_i{(\mu)}S({\cal N}_i[\rho_i{(\mu)}])$. To estimate the lower
bound in Eq.~(\ref{add}), we follow the suggestion of {\cite{holevo}}
and we evaluate ${\cal X}(p_i{(\mu)},\rho_i{(\mu)})$ for the $i$th
mode using coherent states $\rho_i{(\mu)}=|\mu\rangle_i\langle\mu|$
weighted with Gaussian probability distribution
$p_i{(\mu)}=\exp[-|\mu|^2/N_i]/(\pi N_i)$, $N_i$ being the average
number of photons of the mode. In this case, Eq.~(\ref{add}) becomes
$C\geqslant\max_{N_j}\sum_{i}g(\eta_iN_i)$, where again the maximum
must be taken under the average energy constraint (\ref{energ}). The
corresponding Lagrange equation has solution given by Eq.~(\ref{ssss})
with $\zeta=1/\sqrt{\eta}$, so that $C\geqslant{\cal
  T}\sqrt{\eta}\;R_C$ {\cite{paper2}} (see Fig.~\ref{f:numeric}b).
Notice that for $\eta=1$ the equality holds, since the noiseless
channel is known to be additive and we reobtain the results of
{\cite{caves}}.  A closed expression for $Q$ is also not known.
However, for $\eta\leqslant 1/2$ the no-cloning theorem can be used to
show that $Q=0$, as in the case of the erasure channel
{\cite{erasure,paper2}}. For $\eta>1/2$, a lower bound can be obtained
evaluating the coherent information $J({\cal N},\varrho)=S({\cal
  N}[\varrho])-S(({\cal N}\otimes\openone)[\Phi_\varrho])$ on
unentangled non-squeezed Gaussian inputs {\cite{preskill,paper2}}. In
fact, random quantum codes can send quantum information down a noisy
channel at a rate given by the coherent information {\cite{seth}}.  In
Fig.~\ref{f:numeric}b this bound is plotted by solving numerically the
corresponding Lagrange equation, which maximizes the expression
\begin{eqnarray}
Q\geqslant\max_{N_j}\sum_ig(\eta N_i)-g((1-\eta) N_i)
\;\labell{coh}.
\end{eqnarray}

{\em Conclusions.---} Up to now only few realistic channels have been
analyzed at the quantum level.  In this paper we studied the wideband
bosonic channel with loss, calculating the entanglement assisted
capacities $C_E$ and $Q_E$ and we gave upper and lower bounds on the
classical and quantum capacities of this channel. The capacity $C_E$
was shown to scale with the square root of the input power as shown
previously for the classical capacities in the noiseless case.
Moreover, we saw that the superdense coding effect allows the sender
to increase the information transferred above the entropy of the input
state if the quantum efficiency is $\eta>1/2$.

This work was funded by the ARDA, NRO, NSF, and by ARO under a MURI
program.

\begin{figure}[hbt]
\begin{center}
\epsfxsize=.9
\hsize\leavevmode\epsffile{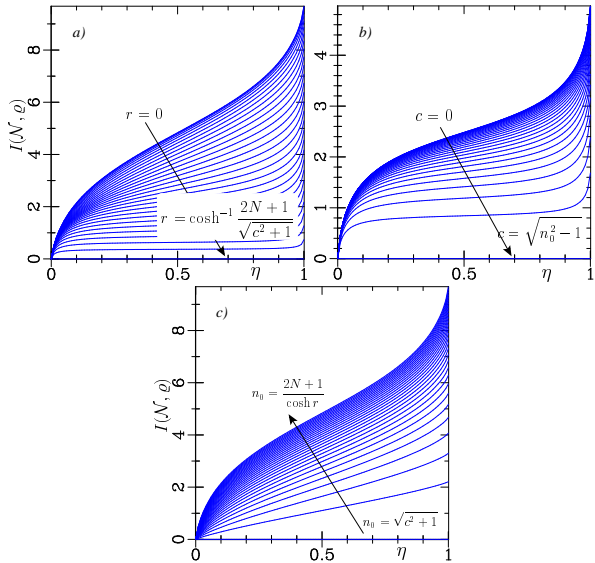}
\end{center}
\vspace{-.5cm}
\caption{Plots of the quantum mutual information $I({\cal N},\varrho)$
  of Eq.~(\ref{ll}): {\it a)} $I({\cal N},\varrho)$ decreases with $r$
  (here $c=0,\;m=0$); {\it b)} $I({\cal N},\varrho)$ decreases with
  $c$ (here $r=0,\;m=0$); {\it c)} $I({\cal N},\varrho)$ increases
  with $n_0$, i.e. decreases with $m$ (here $r=0,\;c=0$). In all plots
  $N=10$. } \labell{f:grafico}\end{figure}

{\em Appendix.---}
In {\cite{holevo}} it has been shown that, for a given value of the
correlation matrix $\alpha$, the quantum mutual information $I({\cal
  N},\varrho)$ for a single mode $a$ achieves its maximum value on the
Gaussian state
\begin{eqnarray}
\varrho=\frac\hbar{2\pi}\int dz\;\exp\left[-i(\Delta q,\Delta
p)\cdot z^T- z\cdot\alpha\cdot z^T/2\right]
\;\labell{gaus},
\end{eqnarray}
where $z$ is a real bidimensional line vector and $q$ and $p$ the two
orthogonal quadratures $q=\sqrt{\hbar/2}(a+a^\dag)$,
$p=-i\sqrt{\hbar/2}(a-a^\dag)$. In order to evaluate the effect of the
squeezing on the quantum mutual information of the single mode
channel, it is convenient to introduce the following parametrization
for the correlation matrix $\alpha$:
\begin{eqnarray} \alpha=\frac\hbar 2 \left[\matrix{n_0e^r&c\cr
      c&n_0e^{-r}}\right] \;\labell{nuovoalpha},
\end{eqnarray}
where $r$ is the squeezing parameter. These parameters are related
through the average number of photons $N$ by the conditions
$\sqrt{c^2+1}\leqslant n_0=[(2N+1)-m]/\cosh r$: the first relation
derives from the strong version of the uncertainty relation, while the
second from the average energy constraint (with $m=\langle
q/\hbar\rangle^2+\langle p/\hbar\rangle^2$). With these definitions
the quantum mutual information becomes \begin{eqnarray} &&I({\cal
    N},\varrho)=g(\gamma(1))+g(\gamma(\eta))-g(\gamma(1-\eta))
  \;\labell{ll}, \\
  &&\gamma(\eta)\equiv\sqrt{\left[\eta\lambda_++\frac{1-\eta}2\right]\left[
      \eta\lambda_-+\frac{1-\eta}2\right]}-\frac 12
  \labell{defdilambda1},
\end{eqnarray}  
where $\lambda_\pm=\frac 12[n_0\cosh(r)\pm\sqrt{(n_0\sinh(r))^2+c^2}]$
are the two eigenvalues of $\alpha/\hbar$.  Notice that for $\eta=1$,
when all the photons reach the receiver, $I({\cal N},\varrho)$ is
twice the entropy of the initial state, as predicted by the superdense
coding effect {\cite{sdc}}. In general, one can verify that   $I({\cal
N},\varrho)$ is smaller than the initial entropy for $\eta<1/2$ and
greater for $\eta>1/2$: the effect of superdense coding is, hence,
evident only in this last case. Since the eigenvalues $\lambda_\pm$
are related with the average number of photons $N$ as   \begin{eqnarray}
\lambda_++\lambda_-=2N+1-m \;\labell{en},
\end{eqnarray}
one can show that the maximum of $I({\cal N},\varrho)$ for fixed $N$
is obtained for $\lambda_+=\lambda_-$. This is equivalent to requiring
$r=0$ (i.e. no energy should be ``wasted'' in squeezing the input---
see Fig.~\ref{f:grafico}a) and $c=0$ (see Fig.~\ref{f:grafico}b). This
last condition attests that the best one can do to convey information
is to send maximally mixed states, since the parameter $|c|$ measures
the purity of the initial state.  Choosing the maximum value of $c$
corresponds to sending a single pure state and conveys no information.
Finally, since $I({\cal N},\varrho)$ is an increasing function of
$n_0$, it can be further maximized by choosing $n_0=2N+1$ (i.e.  its
maximum allowed value achieved when $\langle q\rangle=\langle
p\rangle=0$--- see Fig.~\ref{f:grafico}c).  With this choice,
Eq.~(\ref{defdilambda1}) becomes $\gamma_{opt}(\eta)=\eta\;N$, which
maximizes the quantum mutual information as
\begin{eqnarray}
c_E({\cal N},\varrho)&\equiv&\max_{\varrho\;|\;\langle a^\dag a\rangle=N}
I({\cal N},\varrho)\nonumber\\
&=&
g(N)+g(\eta\;N)-g((1-\eta)N)
\;\labell{ccc},
\end{eqnarray}
as reported in Eq.~(\ref{cedii}).


\begin{references}
\bibitem{notaeta} For $\eta=0,1$ one has to apply the Lagrange
  procedure directly to the functions $c_E(N_i,0)$ and $c_E(N_i,1)$.
  This gives the same result one would get extending by continuity
  the solution of Eq.~(\ref{lagr1}) under the constraint
  (\ref{energ}).
\bibitem{nota1} Notice that, since we have replaced the summations
with the integrals, Eq.~(\ref{cidiefin}) is valid up to corrections of
order $1/{\cal T}$.
\end{references}
\end{document}